\documentclass[12pt]{article}

\usepackage[normalem]{ulem}
\usepackage{amsmath}
\usepackage{amsthm}
\usepackage{amsfonts}
\usepackage{amssymb}
\usepackage{comment}
\usepackage{xcolor}
\usepackage{times}
\usepackage{graphicx}
\usepackage{float}
\UseRawInputEncoding

\newcommand{\AAu}[1]{\textcolor{black}{#1}}
\newenvironment{sciabstract}{%
\begin{quote} \bf}
{\end{quote}}

\newcommand{\HO}{H$_2$O$_2 \,$}

\newcounter{lastnote}

\title{Quantitative measurements of non-equilibrium interactions of catalytic microswimmers with dual colloidal tracers}

\author
{C. Carrasco$^1$, Q. Martinet$^2$, Z. Shen$^{3}$, J. Lintuvuori$^4$, J. Palacci$^2$, and A. Aubret$^{4,\ast}$\\
\\
\normalsize{$^{1}$Department of Physics, University of California San Diego}\\
\normalsize{$^{2}$IST Austria, Klosterneuburg, Austria}\\
\normalsize{$^{3}$ State Key Laboratory for Turbulence and Complex Systems,}\\
\normalsize{Department of Mechanics and Engineering Science, College of Engineering,}\\
\normalsize{Peking University, Beijing 100871, PR China}\\
\normalsize{$^{4}$CNRS, LOMA, UMR 5798, F-33400 Talence, France}\\
\\
\normalsize{$^\ast$To whom correspondence should be addressed; E-mail:  antoine.aubret@u-bordeaux.fr}
}

\date{}

\begin{document} 


\baselineskip24pt


\maketitle


\begin{sciabstract}
Catalytic microswimmers convert the chemical energy of a fuel into motion, sustaining spatial chemical gradients and fluid flows that drive their propulsion. This leads to unconventional individual behavior and the emergence of collective dynamics, absent in equilibrium. The characterization of the non-equilibrium interactions driven by those concentration gradients and flows around microswimmers is challenging owing to the importance of fluctuations at the microscale. Previous experiments have focused on large Janus microspheres attached to a surface, and did not investigate non-equilibrium interactions for freely moving microswimmers of various shapes. Here we show a massive dependence of the non-equilibrium interactions on the shape of small catalytic microswimmers. We perform \AAu{tracking} experiments at high troughput \AAu{to map non-equilibrium interactions between swimmers and colloidal tracers in 2D, accurate down to tracer velocity of $\sim 100nm/s$}. In addition, we devise a novel experimental method combining two types of tracers with differing phoretic mobility to disentangle phoretic interactions in concentration gradients from hydrodynamic flows.  We benchmark the method with experiments on a single chemically active site and on a catalytic microswimmer tethered to a surface. We further investigate the activity-driven interactions of freely moving catalytic dimers as microswimmers, for a wide range of aspect ratio between the active and passive part. We confront our results with standard theoretical models of microswimmers near surfaces and show poor agreement, ruling out phoresis as the main interaction for catalytic swimmers. Our findings provide robust quantitative measurements of the non-equilibrium interactions  of catalytic microswimmers of various geometry with their environment. The work notably indicates the need for theoretical development, and lays the groundwork for the quantitative description of collective behavior in suspensions of phoretically-driven colloidal suspensions.
\end{sciabstract}

\section*{Introduction}
Self-propulsion of catalytic microswimmers originates from the energy transduction of a chemical fuel into work. Microswimmers consume fuel, sustaining spatio-temporal concentration gradients, transduced into hydrodynamic flow by diffusiophoresis (migration of a colloid in the gradient) and osmosis (flow along a surface). This enables catalytic microswimmers to move autonomously. The effect of the spatial-temporal disturbance (concentration and flows) induced by the microswimmers on their environment or by other microswimmers controls the unconventional individual and collective behaviors observed in those non-equilibrium systems. Effectively, catalytic microswimmers constitute the workhorse of active matter and nanorobotics, and the characterization of their non-equilibrium interactions is key to further their applicability \AAu{in those fields}\cite{Bechinger_Volpe-RevModPhys-2016}. The quantitative measurement of the non-equilbrium interactions of microswimmers is however challenging, owing to their weak amplitude and the small scale, imposing a highly fluctuating environment. This situation is reminiscent of the experimental {\it tour de force} that constituted the measurement of the hydrodynamic flow exterted by a swimming  {\it E. Coli} bacteria, which size ($\sim 1$ $\mu$m) and speed compares with current catalytic microswimmers \cite{Drescher_Goldstein-PNAS-2011}. However, catalytic microswimmers pose unique challenges: owing to their propulsion scheme, colloidal tracers respond simultaneously to both the chemical gradients (by diffusiophoresis) and the hydrodynamic flows (from propulsion or osmosis). This can have profound effect on the observed phenomenology depending on the size, surface properties, and geometry of the microswimmers, which affect the relative strengths of these interactions \cite{Boniface_Tierno-NatComm-2024}, and make their characterization difficult. There is a need to disentangle the diffusiophoretic and hydrodynamic components of the migration \AAu{of tracers} in order to achieve quantitative measurement of the non-equilibrium interactions of catalytic microswimmers. \\
To date, quantitative studies of freely moving synthetic microswimmers are limited to much larger systems such as droplets \cite{Blois_Dauchot-PRF-2019}. In addition, investigations on smaller, micron-size catalytic microswimmers are scarce and limited to symmetric (half active, half passive) and spherical Janus particles, \cite{Katuri_Sanchez-ScienceAdvances-2021,Campbell_Golestanian-NatureComm-2019}. Campbell \textit{et al.} suggested that phoresis can be neglected for half-coated Pt microswimmers, with hydrodynamic interactions induced by self-electrophoresis \cite{Campbell_Golestanian-NatureComm-2019}, a conclusion that contrasted with experiments by others with similar swimmers  \cite{Katuri_Sanchez-ScienceAdvances-2021}. Katuri {\it et al.} highlighted the importance of osmotic flows sustained by the confining boundaries, under concentration gradients induced by the catalytic swimmer. The osmotic flows are key to observed behavior of microswimmers near surfaces, whether propelled \AAu{through concentration gradients \cite{Simmchen_Sanchez-NatureComm-2016} or temperature gradients \cite{Bregulla_Cichos-JCP-2019}}. They lead to complex non-equilibrium interactions, dependent on the specific particle configuration of the microswimmers \cite{Katuri_Sanchez-ScienceAdvances-2021} \AAu{with respect to nearby surfaces} and dramatically impact the response of tracer particles in a concentration gradient \cite{Boniface_Tierno-NatComm-2024}. Critically, the relative contributions of osmosis, phoresis, and hydrodynamic flow have been for now quantified primarily with microswimmers tethered to a surface, a situation  that leads to fundamentally different and stronger interactions than for a force-free untethered microswimmer. The experiments were also performed using large ($1 - 2$ $\mu m$) tracers forming packed rings around the microswimmers, limiting the spatial resolution of the observations and potentially distorting the chemical and hydrodynamical flow, as noted by the authors \cite{Katuri_Sanchez-ScienceAdvances-2021}. Such effect cannot be overlooked in the light of the non-reciprocal dynamics later observed with pairs or active and passive particles \cite{Singh_Mark_AdvancedMaterials-2017,Yu_Fischer-ChemComm-2018,Xang_Simmchen-CondensedMatter-2019}. 

In the present work, we provide quantitative measurements of the non-equilibrium interactions of freely swimming catalytic microswimmers with varying geometry. To this end, we synthesize a collection of small colloidal dimers (size $1 - 4$ $\mu$m), composed of a photocatalytic hematite cube of fixed size ($800$ nm) attached to a polymeric sphere, whose radius we vary [Fig.1A-E]. This allows us to compare the non-equilibrium interactions between dimers of varying aspects ratio, with all other physical parameters identical. The non-equilibrium interactions are obtained by quantifying the dynamics of small ($700$ nm) tracers, that exhibit important thermal motion. Experiments are realized in a dilute regime, preventing the formation of packed structures around the microswimmers, which could impact the dynamics \cite{Katuri_Sanchez-ScienceAdvances-2021}. \AAu{We acquire} large collection of data ($\sim 10^7$ frames for each experimental conditions) and perform a thorough statistical analysis of the velocity field, allowing us to measure tracer velocities down to $\sim 1$ \% of the swimmers speed and as low as $100$ nm/s [Fig.1K-O]. Our \AAu{design and analysis} are complemented by a novel approach using two sets of colloidal tracers identical in size and density but differing in surface properties [Fig.2]. It results in tracers particles, for which the \AAu{sensitivity to flows} is unchanged while exhibiting differing phoretic migration in a concentration gradient. We leverage this property to disentangle the phoretic and hydrodynamic contributions of the non-equilibrium interactions. Our approach is first quantitatively validated upon conventional experiments of migration of colloidal tracers around a sink of hydrogen peroxide fuel and of a tethered microswimmer, \AAu{allowing us to extract the relevant experimental parameters}. We then turn to the experimental characterization of the non-equilibrium interactions of untethered catalytic swimmers and show a massive dependency of the non-equilibrium interactions upon the aspect ratio between the active and passive parts. \AAu{We further demonstrate that phoresis, while present, is not dominant.} \AAu{Complex flows arise with microswimmers with large passive components, which are qualitatively captured with a minimal model based on the superposition of flow singularities near a wall and radial phoretic migration. At lower aspect ratio, we observe a dominant, isotropic flow advection, which highlights the specific role of boundaries on concentration gradients and osmotic flows. Our results show a degree of complexity unaccounted for previously and, we are hopeful, will stimulate a critical and fruitful inspection of  current theoretical and numerical models.}

\section{Experimental strategy}
We describe below our experimental strategy and the experimental components required to provide robust and quantitative measurements of the non-equilibrium interactions of catalytic microswimmers.

\subsection{Synthesis of colloidal dimers with varying aspect ratio}
We consider colloidal dimers formed of a photocatalytic hematite (iron-oxide $\alpha$-Fe2O3) bound to an inert inert polymer sphere (3-(Trimethoxysilyl)propyl methacrylate, TPM). The dimers are catalytic microswimmers. They are activated under blue-green illumination, where hematite degrades hydrogen peroxide \HO into water and oxygen. This ultimately yields to concentrations gradients and motion by self-phoresis \cite{Golestanian_Adjari-NewJournalofPhysics-2007}. In order to study the effect of the geometry of the active particle onto the non-equilibrium interactions, we synthesize a collection of dimers, with varying aspect ratios [Fig.1A-E]. We perform all synthesis from a single batch of hematite seeds, preventing variability in the photocatalytic properties of hematite from batch to batch, and allowing us to measure differences rooted in geometrical difference between dimers only. In effect, the size of the hematite is kept constant at $\sim 800$ nm and the inert part is varied by adjusting the amount of TPM oil during the synthesis. We cover a broad of aspect ratio,  $\chi = R_{\text{p}}/R_{\text{a}}$, where $R_{\text{p}}$ is the radius of the passive part, and $R_{\text{a}}$ the average radius of the active part, ranging from $0.9$ for nearly symmetric dimers to $3.5$ for the microswimmers made of a hematite component bound to a large passive sphere [Fig.1A-E]. \\
Experiments are performed in a glass capillary, filled with 6\% vol. \HO fuel in water and pH $\sim$ 6.5, for which the photocatalytic hematite part trails the microswimmer. Observations of the tracers and motion of the swimmers are carried with a high N.A objective (NA=1.45) near the bottom-plane of the capillary, due to the sedimentation of the microparticles. In the absence of illumination, the particles undergo Brownian motion and do not self-propel. When illuminated by blue light (intensity $\sim 1$ $\mu$W/$\mu$m$^2$, $\lambda \approx 450$ nm), the particles self-propel along the surface at typical velocities of $\sim 15$ $\mu$m/s as extracted from the computed Mean Squared Displacement and displacement histograms. All swimmers velocities are comparable, except for the largest aspect ratio, where we found a slower velocity of $\sim 5-10$ $\mu$m/s within our experimental conditions [SI].

\subsection{Experimental observations and statistical analysis}
We study the displacement of the colloidal tracers in areas of typically $30 $ $\mu$m  $\times 30  $ $\mu$m around  microswimmers with varying aspect ratios  $\chi \sim 0.9 - 3.5$ [Fig.1F-O]. Thermal fluctuations of tracers of $700$ nm are visibly important, but enable accurate measurements in the near field of the microswimmers otherwise difficult with repulsive interaction, providing that sufficient statistics is obtained. We work at low density of microswimmers, ensuring that collective interactions between swimmers are negligible, with frames containing more than one swimmer in the field of view removed from analysis. We similarly use low tracer density of $\Phi \sim 10^{-2}$ part.$/\mu$m$^2$,  to prevent tracers collision or jamming at the front of the swimmers, contrary to previous experiments \cite{Katuri_Sanchez-ScienceAdvances-2021}.  The tracer trajectories are reconstructed in the lab frame, readjusted on the instantaneous swimmer orientation and centered on the catalytic component of the microswimmer. We typically record $\sim 10^7$ displacements of tracers at a rate of 40 fps for each considered geometry of the photocatalytic microswimmer. This high-throughput  is required to achieve satisfactory signal to noise ratio [Fig.1K-O]. To highlight this massive experimental undertaking and statistical analysis, we overlay to our measurements of displacements of tracers our signal-to-noise ratio of specific velocity vector $\langle V_{\text{x,y}} \rangle$ $SNR = (\langle V_\text{x} \rangle^2 + \langle V_\text{y}\rangle ^2)\sqrt{N^2}/(\sigma_\text{x}^2 + \sigma_\text{y}^2)$, where the average relative standard deviation of the local velocity $\sigma_{\text{x,y}}$ is reduced by the bin statistics $N$ [Methods].
The resulting velocity plots are shown in [Fig.1K-O]. We show good SNR at typical distances of up to 10 $\mu$m, with tracers velocities being accurately determined down to $\sim 1$\% of the swimmer speed, i.e. $\sim 100$ nm/s. In addition, we unambiguously observe a massive dependence of the non-equilibrium interactions with the aspect ratio. To assess the physical origin of the velocity fields, we further acquire the data with dual tracers that differ in their response to chemical disturbances, as we discuss now.

\subsection{Designing dual tracers to disentangle phoretic and hydrodynamic contributions}
\label{sec:tracers}
Photocalytic swimmers sustain concentration gradients, which, in turn, induce phoretic flows in the medium and osmotic flows on the neighboring surfaces. Colloidal tracers present in the solution and used to measure non-equilibrium interactions can displace phoretically by following concentration gradients \AAu{and be hydrodynamically advected at the same time}. A key challenge is to disentangle the two contributions from the observed dynamics of colloidal tracers. It is noteworthy to realize that the hydrodynamic advection is set by the radius and density of the tracers while phoretic interactions, due to their interfacial origin, are driven by their surface property, e.g. surface charge, which can be independently controlled.   \\
To this end, we synthesize bare silica spheres with diameter $\Phi \sim 700$ nm. The particles are negatively charged with a measured zeta potential $\zeta_{0} \sim -40$ mV under our experimental conditions. They constitute a first set of colloidal tracers. A collection of those particles is subsequently modified and grafted with a primarily neutral polymer brush of Poly(N-isopropylacrylamide) (pNIPAM) to form our second set of tracers \cite{Karg_Hellweg-ChemPhysChem-2006} [Methods \& Fig.2A]. Note that while pNIPAM is a thermo-responsive polymer, it is only considered in this work at constant room temperature for which the polymer brush is extended. Effectively, the polymer brush dramatically increases the viscous drag over the thickness of the polymer brush, \AAu{$\sim 50$ nm from electron microscopy images and light scattering measurements} [SI]. This affects the phoretic mobility \cite{Youssef_Palacci-SoftMatter-2020}, as confirmed by the measured 20-fold reduction of the electrophoretic mobility of the silica particles grafted with pNIPAM compared to the bare silica tracers [SI]. The polymer brush is however thin enough to avoid significant modification of the hydrodynamic radius or density of the particle, thereby the two tracers are considered hydrodynamically equivalent for our level of precision, as independently confirmed also by measurements of their Brownian diffusivity [SI]. In brief, we devised a set of 2 colloidal tracers, with equivalent hydrodynamical behavior but with broadly different phoretic response [Fig.2A]. \AAu{We note that the effect of a polymer brush on interfacial flows depends on the type of imposed gradients and is not affected in the same manner depending on the phoretic mechanism at work \cite{Bregulla_Cichos-JCP-2019,Youssef_Palacci-SoftMatter-2020}}. We thus characterize below the difference of response of the two tracers in the concentration gradient obtained by a chemical sink [Fig.2]. We later leverage the properties of those particles to characterize the non-equilibrium interactions.

\section{Experimental results}

\subsection{Tethered hematite}
We first conduct experiments to characterize the non-equilibrium interactions experienced by the colloidal tracers in a concentration gradient. To this end, we consider a photocatalytic hematite tethered to the bottom substrate of the glass capillary and immersed in a solution of \HO fuel. The hematite is activated by uniform light, inducing a non-uniform radial concentration $c\propto 1/r$, solution of the stationary solution of the diffusion equation $\Delta c=0$. Colloidal tracers respond to this gradient and displace, as a result of flows or phoretic migration. They eventually reach a steady distribution \AAu{under continuous illumination}, where diffusive Fick's flux balances non-equilibrium fluxes [Fig.2B]. Bare silica tracers are noticeably more repelled than the coated tracers forming a distinctly larger prohibited zone away from the hematite. We quantify the velocity field of the bare and pNIPAM-coated silica tracers by tracking and measuring the radial velocity $V^h_{B}$ and $V^h_{C}$, respectively, as a function of their distance $r$ from the center of the hematite [Fig.2C]. \AAu{We periodically switch off the excitation to let enough time for the tracers to diffuse back close to the hematite and increase statistics.} The measurements of the velocity field for the tracers confirm the qualitative assessment made with the density profiles of the tracers and show a radial repulsion away from the hematite for the bare tracers [Fig.2B,C-blue symbols]. In contrast, pNIPAM-coated tracers show vanishing repulsion except in the close vicinity of the hematite [Fig.2B,C-red symbols]. The marked difference in the two datasets confirms the effective reduced mobility of the coated particles vs. bare ones. The quality and accuracy of the measurements allow us to measure a reversal of sign and near field attraction with a small inward radial flow, $V_C^h \sim 100-200$ nm/s for the pNIPAM coated tracers [Fig.2C-inset]. This change of direction of the velocity field is unexpected in a purely \AAu{phoretically repulsive} scenario  but confirmed by the local increase observed in the density profile of tracers ([Fig.2B]). We performed additional experiments in inverted configuration, where hematite are tethered to the upper wall on the capillary, in order to investigate gravitational effect as \HO is denser than water [SI]. The results were unchanged, discounting catalytically-induced density gradients, or thermogravitational convection as origin of the reversal of sign. We intuitively attribute this sign reversal to complex osmotic flows on the substrate of the capillary wall, as recently reported \cite{Katuri_Sanchez-ScienceAdvances-2021, Boniface_Tierno-NatComm-2024}, and confirmed by our model below.
\\
To gain further insight and eliminate the complexity arising from those complex osmotic flows, we take advantage of our set of dual tracers to measure non-equilibrium interactions. As discussed previously in sec.\ref{sec:tracers}, we designed colloidal tracers, which are hydrodynamically equivalent but phoretically different. As  such, we expect that all flows, including the osmotic flows generated by the surface, will be removed by inspecting the difference of velocity field $\delta V^h=V^h_B-V^h_C$ between the bare and the coated tracers.  We perform this subtraction and show a monotonic decay with a $\sim 1/r^2$ dependency, consistent with a phoretic repulsion induced primarily by neutral diffusiophoresis, i.e. with a velocity $\propto -D_{\text{B,C}} \nabla_r c(r)$, with $c(r) = c_0/r$ [Fig.2D]. The fit of the data provides the difference of phoretic \AAu{repulsion strengths} between the bare and coated tracers $(D_{\text{B}}-D_{\text{C}})c_0 = 8 \pm 2$ $\mu$m$^3$/s. 
To further our physical understanding of the effect of osmotic flows in the experiment, we perform numerical calculations of the flow field generated by a spherical particle tethered on a surface with homogeneous reaction rate on its shell [Fig.2C,E] [SI]. We notably make the assumption that the glass substrate, the bare silica tracers and the catalytic hematite have equal diffusiophoretic mobility, as their surface has comparable surface chemistry.  Following our experimental observation, we compute a slip velocity of the form: $v_s = -D_B\nabla_s c$, on both the catalytic and substrate surfaces, superimposed to a repulsive phoretic migration of the tracer $v = -D_{B,C} \nabla c$, thus accounting for osmotic flows and phoretic migration. This approximate model reproduces  our experimental data remarkably well, after fitting the amplitude of the velocity $V^h_{B}(r)$ for the bare particles, and the ratio of bare to coated phoretic mobilities, $D_{B}/D_{C} = 2 \pm 0.2 $. It remarkably captures the reversal of sign for the interaction for the coated particles and confirms it originates from osmotic flows. Importantly, this framework allows us to quantitatively evaluate the absolute value of phoretic contributions from the knowledge of $\Delta V^h$, finding in this case $D_B = 16$ $\mu$m$^3$/s and $D_C=8$ $\mu$m$^3$/s. 
In brief, our experimental results with tethered hematite show that colloidal particles are adequate tracers for the quantification of non-equilibrium  interactions provided the superposition of i) osmotic flows at the substrate and catalytic particle surface, and ii) a repulsive phoretic migration. Importantly, the measurements provide a quantitative reference point to model the effects induced by chemical gradients in the upcoming experiments with photocatalytic colloidal dimers of variable aspect ratio $\chi$.

\subsection{Photocatalytic dimers tethered to the capillary substrate}
We now turn to experiments with photocatalytic dimers tethered to the capillary substrate, aiming to characterize and quantify the hydrodynamic flows generated by the propulsion. In this configuration, the microswimmer is not force-free owing to the net force opposing phoretic self-propulsion \cite{Moyses_Grier-SoftMatter-2016, Campbell_Golestanian-NatureComm-2019,Drescher_Goldstein-PNAS-2011,Katuri_Sanchez-ScienceAdvances-2021}. Therefore, it exerts a net force on the fluid, similiar to a force monopole in the far-field and at all values of the aspect ratio $\chi$. We therefore consider a photocatalytic dimer at a single aspect ratio, $\chi = 1.9$ [Fig.1C]. We analyze the 2D velocity field of tracers around tethered swimmers, $\mathbf{V^t_{B,C}}$ in the reference frame of the lab [Fig.3]. \\
For both bare and coated tracers, the velocity field presents a strong inward velocity at the front and outward velocity at the back.  The tethered swimmer effectively pumps fluid along its swimming direction, a characteristic of a point force (Stokeslet) in the plane of observation  [Fig.\ref{Fig3}A]. The lateral displacement of the tracers however differs. While the interaction appears mainly radially repulsive for bare tracers, it exhibits stronger lateral attraction for coated tracers. Once again, we leverage our dual tracers in order to gain further insight in the non-equilibrium interactions. In order to remove the hydrodynamic components (identical for both tracers) and solely focus on the effect of the phoretic migration of the tracers in the concentration field, we compute the difference $\Delta \mathbf{V^t} = \mathbf{V^t_{B}} - \mathbf{V^t_{C}}$  [Fig.\ref{Fig3}B,C]. Remarkably, the velocity field for $\Delta \mathbf{V^t}$ becomes purely isotropic, suppressing the anisotropy observed for individual tracers [Fig.3C]. In addition, the amplitude of the migration shows $1/r^2$ decay, thus confirming that the use of dual tracers and subtraction of the displacement fields allowed us to isolate the phoretic migration in the concentration gradient sustained by the tethered microswimmer. The amplitude of phoretic repulsion is however stronger than for the tethered hematite only $|\Delta \mathbf{V^t}|/|\Delta \mathbf{V^{h}}| \approx 4 \pm 1$, a difference that we attribute to a variation on the surface chemistry of hematite during the synthesis of photocatalytic microswimmers from individual hematite particles [Fig.3C].\\
	We next compare our experimental observation to an elementary model that consists in the superposition of a Stokes monopole with the phoretic migration as measured in the previous section for tracers responding to a tethered hematite, accounting for the scaling of $|\Delta \mathbf{V^t}|/|\Delta \mathbf{V^{h}}|$. For a swimmer bound to the surface, a net force is applied by the substrate to compensate for the otherwise swimming (force-free) body. The force is given by $ f \sim 6\pi l \eta V_{sw}$,  where $l = R_a+R_p$ and $\eta$ the water viscosity at $300$K. Taking the measured average $V_{SW}\approx 15$ $\mu$m/s for a $\chi=1.9$ swimmer, we find $f \approx 0.35 \pm 0.05$ pN. We simulate the flow field of a Stokeslet near an interface at prescribed height $h=R_p \sim 800nm$, for which the analytical expression is known \cite{Spagnolie_Lauga-JFM-2012}. In order to account for the depth of field of the objective, we further average the computed flows from different planes within the depth of field below and above the plane of the swimmer (see [Methods]). The simulated data are presented in [Fig.\ref{Fig3}D] and show a remarkable agreement with the complex velocity fields observed experimentally for both types of tracers and {\it without} any fitting parameter [Fig.3D]. This confirms the potent character of our approach using dual tracers to gain quantitative insight on the non-equilibrium interactions induced by the microswimmers, and to succesfully evaluate phoretic migration against hydrodynamic advection.

\subsection{Freely swimming microswimmers}
We now investigate the displacement field of colloidal tracers following the passage of freely swimming microswimmers of varying aspect ratio $\chi \sim 0.9- 3.5$ [Fig.1]. Experiments  are repeated with bare silica tracers [Top panels] and  pNIPAM-grafted silica spheres, for which the phoretic migration is hindered [Bottom panels]. We represent the velocity field  of the tracers normalized by the instantaneous swimmer velocity to compare experiments  $\mathbf{\tilde{V}^f_{B,C}}=\mathbf{V^f_{B,C}}/V^f_{0}$, $V^f_{B,C}$ where $V^f_{B,C}$ is the instantaneous velocity field of the bare (B), respectively  pNIPAM coated (C) tracers and $V^f_{0}$ the instantaneous swimmer velocity. \AAu{For each swimmer, we have selected the top $\sim 50 \%$ velocities, which correspond to configurations where the swimmer is observed to be mostly parallel to the interface [SI]. This prevents strong variations in the velocity field that could arise from stochastic reorientations of the swimmers.}\\
A few comments are in order following our experimental results.  First, the velocity field of the tracer is massively dependent on the geometry of the active dimer: radial, isotropic repulsion for $\chi = 0.9$ becomes complex and anisotropic with recirculation loops with increasing aspect ratios,  $\chi>3$. Second, it is remarkable that the bare and pNIPAM coated tracers show \AAu{only a modest difference in measurements for all values of $\chi$}, hinting at the (unexpected) predominance of the hydrodynamic advection over phoretic migration for the freely swimming microswimmer. Remarkably, the velocity field of tracers for near symmetric microswimmers, $\chi\leq1.5$, significantly departs from the \AAu{stresslet flow (force dipole)} previously reported for bacteria \cite{Drescher_Goldstein-PNAS-2011} and previously suggested for Pt-coated Janus microswimmers \cite{Campbell_Golestanian-NatureComm-2019}. 

\subsubsection{Evaluation of phoretic migration}
\AAu{We now aim at isolating the effect of the phoretic migration of the tracers induced by the free photocatalytic microswimmer. To this end, we compute, as previously, the difference of the absolute velocity field $\Delta \mathbf{V_\chi^f} = \mathbf{V^f_{B}} - \mathbf{V^f_{C}}$ for each aspect ratio $\chi$.} Remarkably, the measurements collapse for all values of $\chi$ onto the $1/r^2$ decay observed for the tethered microswimmer, \AAu{i.e. $|\Delta \mathbf{V^f}|/|\Delta \mathbf{V^{h}}| \sim 4$, independent on the variability in swimmers velocity with the aspect ratio [Fig.4A \& SI]. This confirms that the phoretic component of the non-equilibrium interactions is quantitatively decoupled from the hydrodynamics, and independent on the aspect ratio}. This constitutes a reasonable outcome as all photocatalytic dimers were synthesized from a single batch of photocatalytic hematite. In addition, the agreement with the result of the tethered microswimmer confirms that the chemical gradient is steady in the moving frame of the swimming particle, as expected given the low Peclet dynamics of the system $Pe=V_0 R_a/D \ll 1$, where D is the diffusion of \HO. In light of these results, we quantitatively evaluate the phoretic contribution from our previous knowledge of $D_B/D_C \sim 2$ [Fig.4A,B].

For the two types of tracers, we compare the isotropic phoretic repulsion to the near isotropic interactions observed at low aspect ratio [Fig.4B]. We find a comparable amplitude between the two in the range $1 - 6$ $\mu$m, however with the experimental data presenting a decay $1/r^d$ with $d \sim 3 - 4$, slightly stronger for coated particles and decaying more rapidly than the $1/r^2$ decay of a purely phoretic origin, as we observed previously. 
As for our previous experiments, this suggests that hydrodynamic fluxes, from osmosis and propulsion, screen phoretic migration in the far field. The independence of the phoretic interactions on the aspect ratio, coupled to a net loss of isotropy in the velocity field upon increasing values of $\chi$, further indicates a change of the relative contributions between hydrodynamics fluxes from propulsion and the effects of chemical gradients (osmosis and phoresis).

\subsubsection{Evaluation of hydrodynamic flows from propulsion}
We demonstrated previously that the use of dual tracers allows us to gain physical insight \AAu{by disentangling the phoretic migration of tracers from the hydrodynamic flows}. \AAu{We now compare our experimental observations for the freely swimming catalytic microswimmers with theoretical and numerical predictions of the hydrodynamic flux generated by propulsion}.
 \\
 A theoretical description of the flow of phoretic dimers in a homogeneous, unbounded medium has been previously proposed \cite{Reigh_Kapral-SoftMatter-2015,Michelin_Lauga-ScientificReports-2017} but there exists to our knowledge no model that describes the hydrodynamic flow of a photocatalytic dimer near an interface. We therefore propose a method to extract the main features of the flow generated by swimming dimers near a non-slip wall by extrapolating solutions near an interface from exact analytical solutions in the bulk. Following Ref.\cite{Reigh_Kapral-SoftMatter-2015}, we first compute the wall-free concentration and flow fields of dimers with aspect ratio representative of our experiments, $\chi \in [0.9 - 4]$ assuming identical phoretic mobilities for both the active and passive parts. We next expand the velocity field in terms of Stokes singularities to account for the influence of the substrate, as exact mirror solutions of arbitrarily oriented singularities are well-established \cite{Spagnolie_Lauga-JFM-2012} [See methods] . 
 
We develop the flow around free swimmers as the sum of the flows from a force dipole $\mathbf{u}_{fd}$, a source dipole $\mathbf{u}_{sd}$, as conventional with the squirmer model, and add a force quadrupole $\mathbf{u}_{fq}$ and a source quadrupole $\mathbf{u}_{sq}$ to account for the shape asymmetry of the dimers \cite{Spagnolie_Lauga-JFM-2012,Ishimoto_Gaffney-PRE-2013}:
\begin{equation}
\textbf{\~{u}}(\textbf{r}) = A \textbf{u}_{fd} (\textbf{r}) + B \textbf{u}_{sd} (\textbf{r})+ C \textbf{u}_{fq} (\textbf{r})+ D \textbf{u}_{sq} (\textbf{r})
\label{eq:Floweq}
\end{equation}

We find that the hydrodynamic flows generated by phoretic dimers are well described in terms of flow singularities, which relative amplitudes strongly depends on the aspect ratio of the microswimmer [SI]. In particular, we observe a a transition from puller to pusher with increasing aspect ratio, in line with previous numerical studies on thermophoretic heterodimers \cite{Wagner_Ripoll-EPJE-2021} [Fig.4C]. The transition occurs at $\chi \sim 2$ [SI], as extracted from the change of sign of the conventional squirming parameter $\beta=-(4/3)A$ \cite{Ishimoto_Gaffney-PRE-2013}. At low aspect ratio, force and source dipoles are enough to model the flow field. For values of $\chi \gtrsim 1.5$, however, the quadrupolar term are significant within our range of observation and cannot be neglected, even if it finally reaches the $1/r^2$ stresslet decay in the far-field [SI].
From the knowledge of the set $\{A,B,C,D\}$, we readily extrapolate the hydrodynamic field near an interface from the mirror images of the singularities. The anisotropy of the flow is conserved for each aspect ratio, but decays faster in the far-field ($\propto 1/r^4$) [SI], and presents a qualitatively different behavior in the near field, with additionnal recirculation loops observed even at low aspect ratio compared to bulk computations [Fig.4C].

\subsection{Comparison of the model with experimental observations}
From our numerical prediction, it is clear that the isotropic interactions observed at the lowest aspect ratio ([Fig.1F,K]) cannot result from the hydrodynamic fluxes generated by the propulsion, which are expected to present a net puller anisotropy, with lateral repulsion and front and backward attraction. Additionnally, while the predicted phoretic migration quantitatively match the measured amplitude in the intermediate field ($\sim 5$ $\mu$m) [Fig.4B], it also cannot explain by itself the fast decay observed, nor the change of sign of the velocity at distance $\sim 8$ $\mu$m for the pNIPAM-coated tracers.

Conversely, the experiments at large aspect ratio qualitatively present the main features of a pusher ([Fig.1J,O]), with lateral attraction and front and backward repulsion, and recirculation loops observed near the front and similar to the ones obtained with our calculations [Fig.4C]. It underlines the increased contribution of hydrodynamic fluxes stemming from propulsion, relative to other non-equilibrium interactions, in particular phoretic repulsion.

To gain more insights on the cumulative effects of non-equilibrium interactions, we first add the computed phoretic contributions to the hydrodynamic velocity field of the tracers originating from propulsion, as predicted from eq.\eqref{eq:Floweq} ([Fig.4C, bottom panels]): $\mathbf{\tilde{u}}^f_{B,C} = D_{B,C}/(V_0 r^2) + \textbf{\~{u}}(\textbf{r})$, disregarding osmosis. The velocity fields predicted for the two extreme aspect ratio $\chi=0.9$ and $\chi=3.5$ are presented on [Fig.4D]. In both cases, it predicts phoretic migration to overcome hydrodynamic fluxes, with isotropic repulsion for all values of $\chi$, in stark contrast to our experimental observations. Our analysis and observations therefore suggest a strong screening of phoretic migration by osmotic and surface flows, induced by the presence of the wall. Consequently, phoretic interactions are not the dominant component of non-equilibriums interactions for microscopic dimers, in contrast to what is widely assumed in models of collective behavior for catalytic microswimmers \cite{Liebchen_Lowen-JCP-2019,Liebchen_Cates-PRL-2017}.

The rightful evaluation of osmotic flows for freely moving particles is challenging, owing to the fundamental difference between stalled and force-free swimmers. In our previous experiments on tethered particles, osmotic flows originated from concentration gradients on the glass substrate and on a stalled hematite. The model accounted for the breaking of symmetry of the concentration field in the vertical direction, and the subsequent slip velocity established on the tethered hematite induced additionnal flows by pumping the fluid. For freely swimming dimers, the hematite is untethered and force-free. Therefore, we show that a simple description of the total velocity field using the osmotic fluxes and phoretic interactions scaled from the experiments on fixed catalytic center ([Fig.2C]) wrongfully models the data, by breaking the no-force condition. Checking on this by adding the scaled osmotic flows in $\mathbf{\tilde{u}}^f_{B,C}$ effectively reduces the effect of phoretic repulsion, but the contribution of chemical gradients remains highly overestimated at both aspect ratio, and the simulated curves show poor agreement with experiments [Fig.4E].

\subsection{Discussion}
Our experimental results show non-equilibrium interactions around free swimmer are only moderatly dependent on the tracers for all aspect ratio. A simple theoretical model highlights that the relative contribution of hydrodynamic flows to the non-equilibrium interactions varies strongly with the aspect ratio. In contrast, phoretic interactions remain constant for all aspect ratio, as prescribed by our synthesis procedure, and do not dominate the map of interactions in our experiments. Our results unveil the presence of strong advective flows that compete with phoresis and flows from translational propulsion, inducing long-range attraction and isotropic interactions at low aspect ratio. Our analysis points the need to rightfully evaluate all surface flows induced by the concentration gradients, accounting for the specific swimmer geometry and state of motion.\\
The dynamics of freely moving particles near a wall emerges from a non-linear coupling between hydrodynamic and chemical interactions with the wall, affecting the propulsion speed, height, and orientation of the particles. The breaking of axysimmetry in the vertical concentration gradient further yields to additionnal flows in the observation plane, as recently reported with experiments on large spherical droplets \cite{Blois_Dauchot-PRF-2019}. For catalytic swimmers, the coupling between hydrodynamic fluxes and concentration near surfaces has been described theoretically, but focused on spherical Janus particles. They were shown to exhibit various behaviors depending on the geometry and surface properties of the microswimmer, such as gliding, sliding, self-trapping, hovering, or scattering on surfaces \cite{Spagolie_Lauga-SoftMatter-2015,Ibrahim_Liverpool-EL_2015,Popescu_Dietrich-EPJE-2018,Bayati_Najafi-SoftMatter-2019,Uspal_Tasinkevych-SoftMatter-2015,Shen_Lintuvuori-EPJE-2018,Liebchen_Mukhopadhyay-JPCM-2022}, with some of these effects reported in experiments \cite{Ketzetzi_Kraft-NatureComm-2022,Simmchen_Sanchez-NatureComm-2016,Das_Ebbens-NatureComm-2015}. However, theoretical developments remain to be undertaken for catalytic swimmers of various size and morphology, such as dimers. While computations from Stokes singularities can give a reasonnable picture of the flow in the far-field, they are generally not accurate to model the full hydrodynamics of solid bodies in the near field or close to geometrical boundaries \cite{RojasPerez_Michelin-JFM-2021,Lauga_Michelin-PRL-2016,Drescher_Tuval-PRL-2010}.

Our studies and analysis emphasize the need for a more accurate description of non-equilibrium interactions for microswimmers near surfaces. The description should systematically entail the effect of the wall on chemical gradients, osmotic flows, and of the geometry of the microswimmers, a near-field effect salient to our experiment but overlooked by current models.
%Besides the effect of the nearby surface, we briefly note that inhomogeneous rate of photocatalysis at the surface of the particles could also originate from anisotropic light absorption of the hematite, induced by the specific geometry of our system where illumination is made from the bottom \cite{Aubret_Palacci-SoftMatter-2018,Chen_Zhang-AdvancedMaterials-2017,He_Lin-Angewandte-2017}. Such detailed description is beyond the scope of this work.\\

\section{Conclusion}
We investigated the velocity field generated by phoretic dimers near surfaces for various aspect ratio, and unveiled a massive dependency of the aspect ratio on the effect of swimmers on their environment. We proposed an experimental approach using dual colloidal tracers with differing surface properties. This method notably allows one to extract phoretic interactions even when entangled with complex osmotic flows. The velocity field of tethered microswimmers is successfully described by considering phoretic repulsion, osmotic flows, and a force monopole. For freely moving dimers, we brought to light additional advective flows, and showed that hydrodynamic dipolar interactions and phoretic interactions are not necessarily dominant.
The work notably indicates the need for theoretical development, and lays the groundwork for the  quantitative description and study of collective behavior of suspensions of phoretically-driven colloidal suspensions. We are hopeful it will stimulate a critical and fruitful development of  current theoretical and numerical models.

\newpage
\begin{figure}[htbp]
\centering
\includegraphics[scale=0.5]{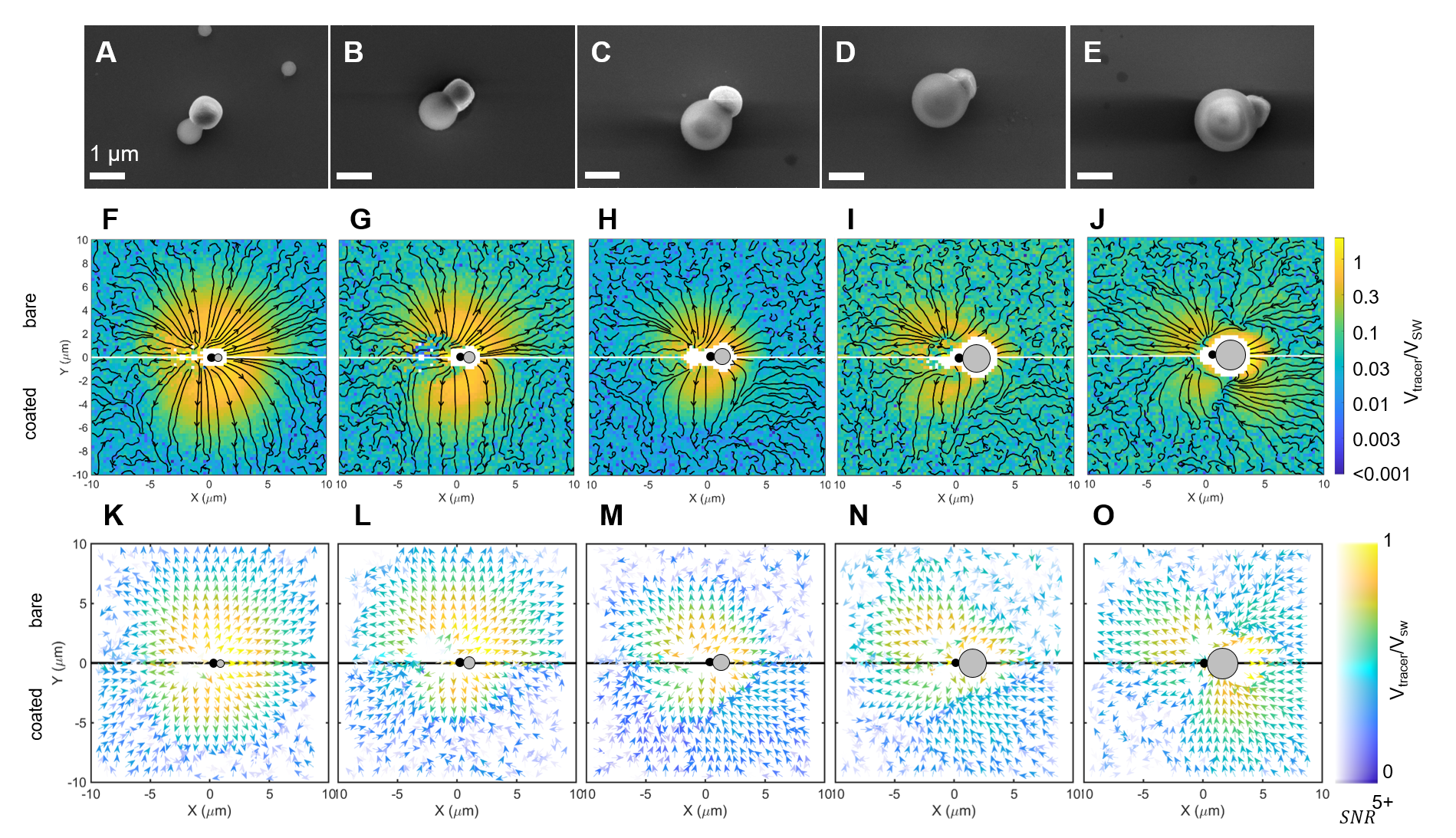}
\caption{\textbf{Direct mapping of the velocity field of passive tracers around swimmers of various aspect ratio}. \textbf{A-E)} Scanning Electron Microscopy images of the five microswimmers used in the study, covering a range of aspect ratio, from left to right : $\chi = 0.9 \pm 0.1$, $\chi = 1.4 \pm 0.2$, $\chi = 1.9 \pm 0.2$, $\chi = 3.2 \pm 0.4$, $\chi = 3.5 \pm 0.6$. \textbf{F-J)} Corresponding maps of the velocity fields of the tracers for bare (top) and pNIPAM-coated (bottom) tracer particles with streamlines, normalized by the instantaneous speed of the swimmer. \textbf{K-O)} Vector plots of the velocity fields represented using a specific color scale that accounts for the statistical accuracy of each measurement. Lightness is adjusted to improve contrast for bins with $SNR \gtrsim 1$, while reducing the perception of noisy data. Accurate measurements of velocities are obtained down to $\sim 1\%$ of the swimmer velocity.}
\label{Fig1}
\end{figure}

\begin{figure}[htbp]
\centering
\includegraphics[scale=0.5]{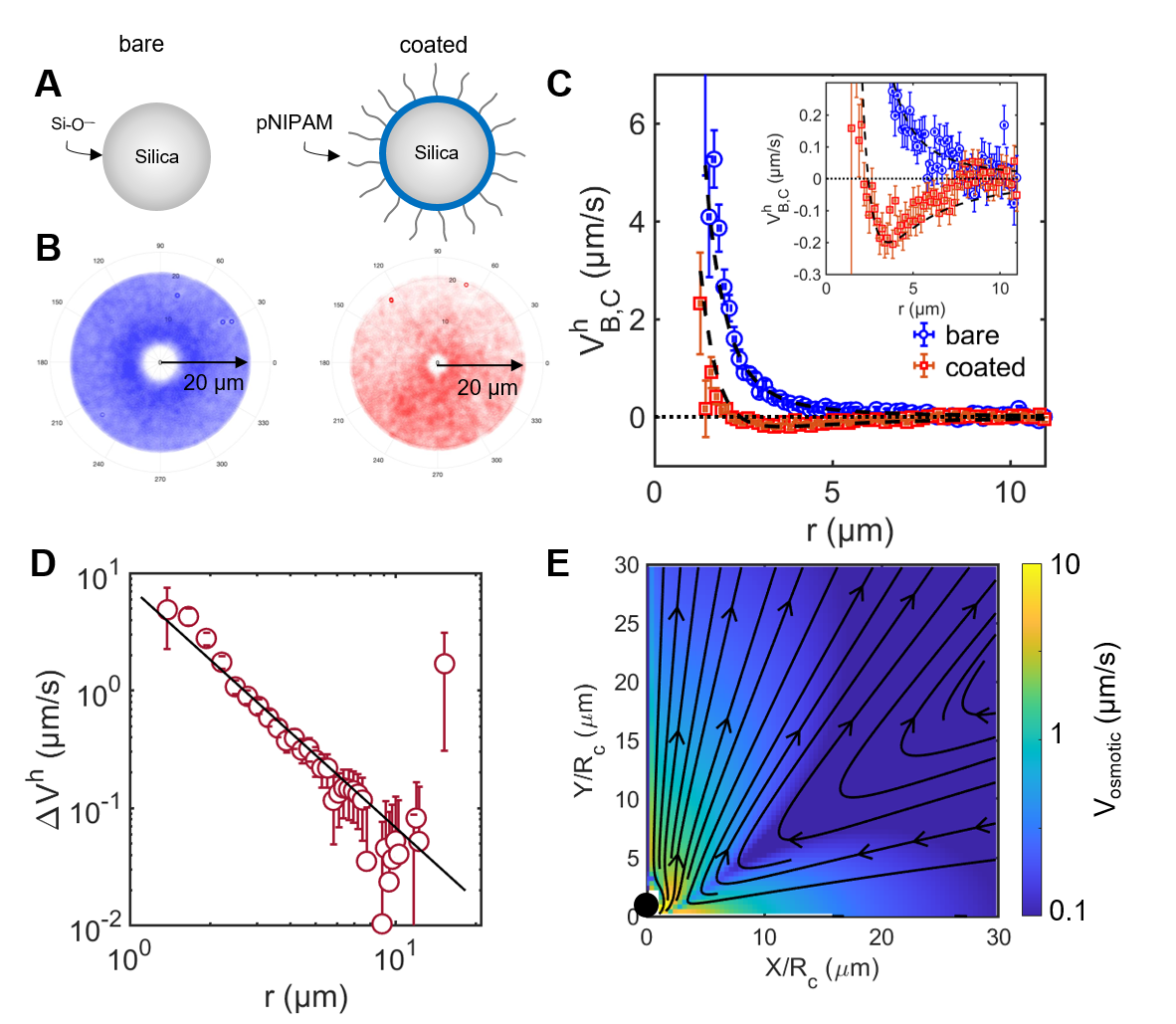}
\caption{\textbf{Response of tracers to a catalytic, active particle tethered on the surface}. \textbf{A)} Design of dual tracers: bare silica, showing $SiO^{-}$ surface groups in water, and pNIPAM coated silica (thickness $\sim 50$ nm). The polymer brush increases the viscous drag in the interfacial layer, and reduces the diffusiophoretic mobility. \textbf{B)} Density profiles of bare (blue, left) and pNIPAM-coated (red, right) tracers around a single hematite particle situated at $r=0$. The density profiles show isotropic repulsion, however reduced for coated tracers compared to the bare ones. \textbf{C)} The angular averaged radial velocity, plotted for bare (blue circles) and coated (red squares) tracers. Dashed lines are simulated data (see \textbf{D)}), obtained for $D_B/D_C = 2$. Dotted line is a guideline for $V=0$. \textit{Inset}: zoomed in portion of the velocity, showing the negative velocity of coated tracers. \textbf{D)} Interpolated difference between the velocities of bare and coated particles. The decay shows agreement with a purely repulsive phoretic interaction decaying as $1/r^2$ (black solid line), with an amplitude of $\sim 8 \pm 2$ $\mu$m$^3$/s, as suggested by neutral diffusiophoresis. \textbf{E)} Computed osmotic flows induced by a catalytic center of homogeneous reaction rate (black disk) of radius $R_c$. The velocity field shows radial attraction near the interface, and outward repulsion far above the catalytic center.}
\label{Fig2}
\end{figure}

\begin{figure}[htbp]
\centering
\includegraphics[scale=0.5]{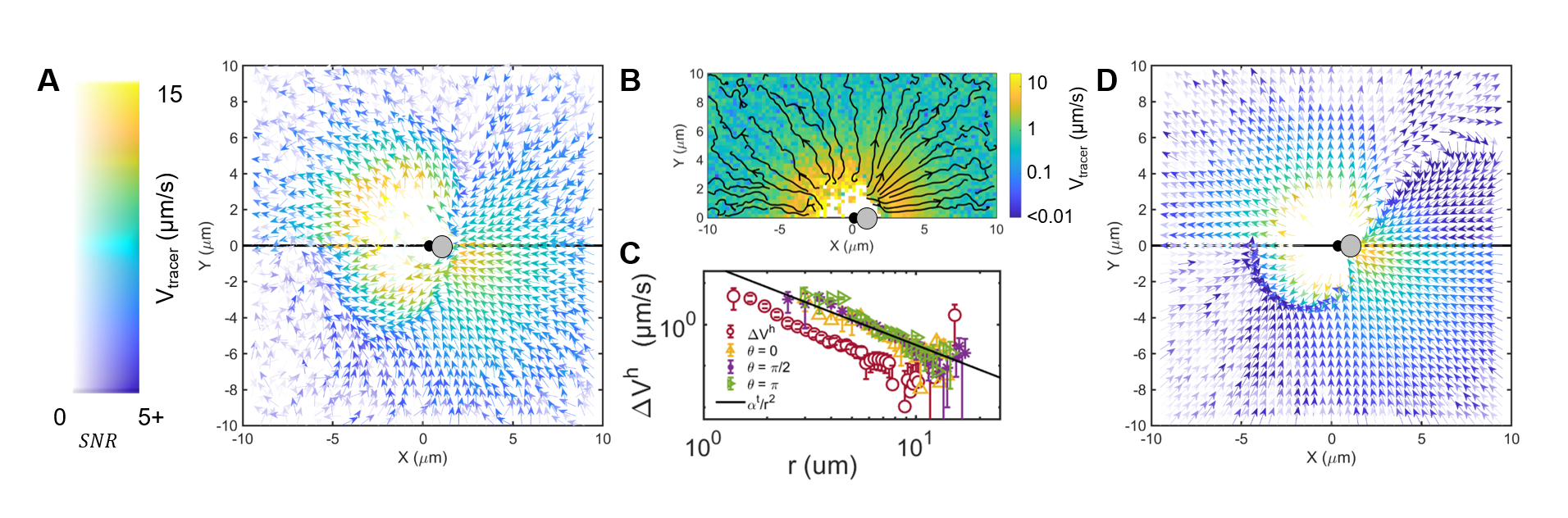}
\caption{\textbf{Response of tracer particles to the presence of a tethered swimmer activated by light}. \textbf{A)} Absolute velocity plots for bare (top) and coated (bottom) particles around a $\chi = 1.9$ dimer. It shows a net attraction at the front, with a repulsive interaction closer to the active part. The color scale is the same as in [Fig.1K-O]. \textbf{B)} Plot of the velocity difference between the bare and coated tracers, showing isotropic repulsion. \textbf{C)} Radial velocity profile for various angles around the swimmer, compared to the data obtained from hematite. The $1/r^2$ is retrieved, and matches the data from single hematite following an appropriate scaling factor of $ (4 \pm 1) \Delta V^h$. \textbf{D)} Simulated velocity plot, considering a point force of strength $6 \pi \eta l V_{SW}$  parallel to the interface at height $h=R_p$, upon which is added the experimentally extracted effect of chemical gradients (see [Fig.2C]), scaled by $4 V^h_{r,B}$ and $4 V^h_{r,C}$. The simple model catches the main features of the non-equilibrium interactions map.}
\label{Fig3}
\end{figure}

%\begin{figure}
%\centering
%\includegraphics[scale=0.7]{Fig4.png}
%\caption{\textbf{Computation of hydrodynamic flows generated by phoretic dimers and desription in terms of flow singularities}. \textbf{A-C)} Computed flow field generated for three typical aspect ratio $\chi=1$ (A), $\chi=2$ (B), and $\chi = 3.5$ (C). The velocity field is obtained by a superposition of Stokes singularities, as fitted from equation \eqref{eq:Floweq} from analytical solutions in the bulk \textit{(top)}, and extrapolated for a microswimmer parallel to an impermeable interface with no-slip boundary condition \textit{(bottom)}. \textbf{D)} Evolution of the extracted squirming parameter $\beta$ as a function of the aspect ratio $\chi$. A transition from puller to pusher is observed at $\chi=2$. \textbf{E)} Evolution of the quadrupolar amplitudes for the force and source quadrupoles. The force quadrupole gives the dominant contribution, and increases in strength as $\chi$ increases, hence the shape asymmetry. \textbf{F)} Decay curves of the velocity field obtained for a $\chi=1$ swimmer near an interface. The flow is mainly dipolar, being attractive at the front and back, and repulsive on the side of thee swimmer. \textit{Inset:} log-plot of the absolute value of the decay curve, showing the fast $\propto 1/r^4$ decay expected for the lower dipolar mode.}
%\label{Fig4}
%\end{figure}

\begin{figure}
\centering
\includegraphics[scale=0.6]{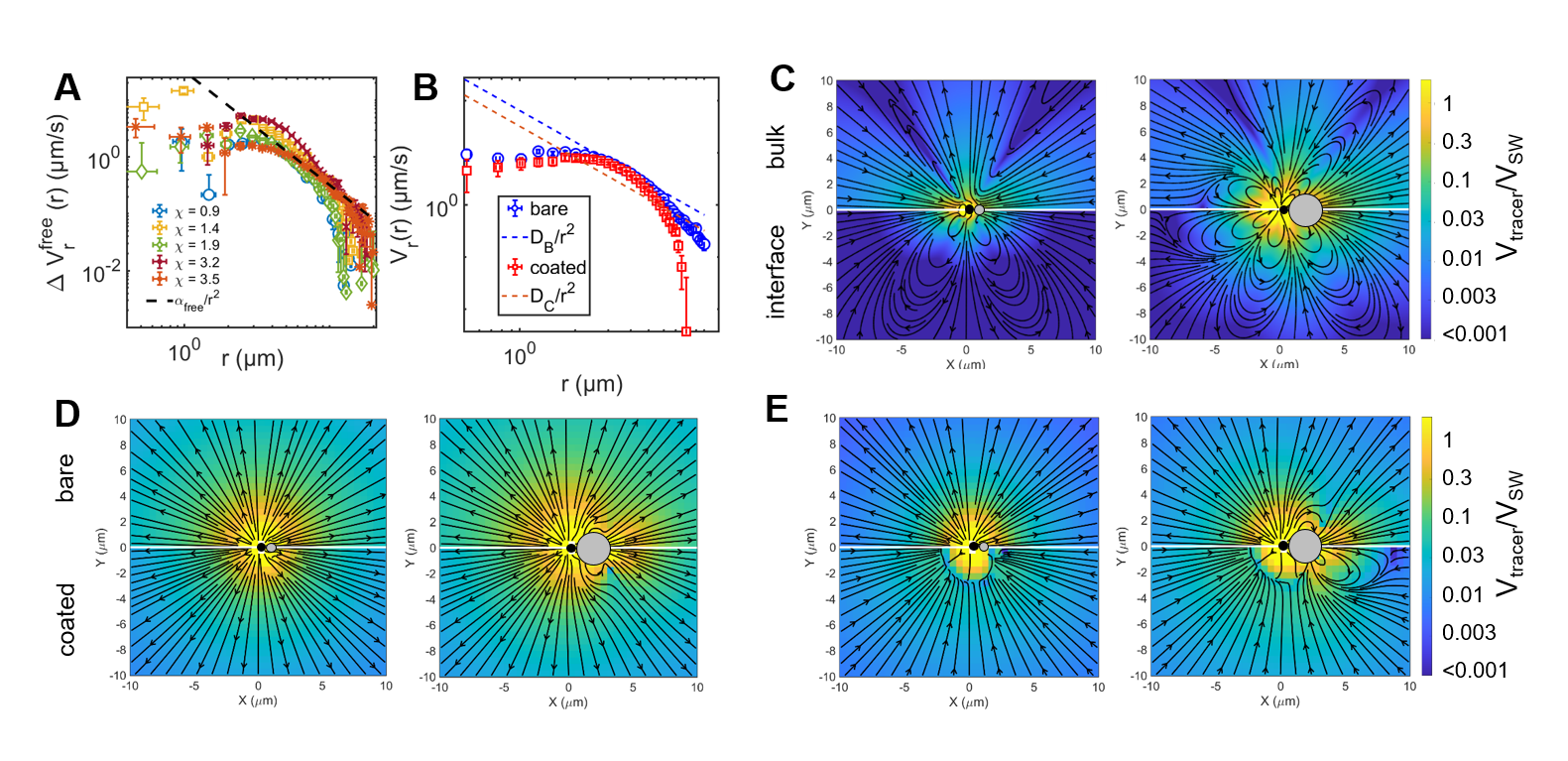}
\caption{\textbf{Disentangling non-equilibrium interactions around freely swimming phoretic microswimmers}. \textbf{A)} Angular average of the absolute velocity plot differentials between bare and coated particles for all aspect ratio between $\chi=0.9$ and $\chi=3.5$. The reasonable collapse of all curves indicates a similar phoretic repulsion for all swimmers, with $1/r^2$ decay (black dashed line) and $|\Delta \mathbf{V^f}|/|\Delta \mathbf{V^h}| \sim 4$. \textbf{B)} Raw angular-averaged experimental decays for the bare (blue circles) and coated tracers (red squares) for $\chi=0.9$, compared to the estimation of phoretic contribution (dashed lines). Phoresis fails to reproduce the observed decay. \textbf{C)} Numerical calculations of the hydrodynamic flow field only, originating from the propulsion of the dimer, as computed from eq.\eqref{eq:Floweq}, for the two extreme aspect ratio $\chi=0.9$ and $\chi=3.5$. The field is computed in bulk from analytical solutions (top) and for the solution extrapolated near an interface, using a sum of singularities (bottom). It shows the stresslet field of a puller (resp. pusher) at low aspect ratio (resp. high aspect ratio). \textbf{D)} Hydrodynamic field as computed in C), on which phoresis is added (no osmosis), for bare (top) and coated (bottom) tracers. For both aspect ratio, it shows a strong dominance of phoretic migration in disagreement with our experimental data [Fig.1]. \textbf{E)} Velocity field as in \textbf{D}), on which the osmotic flows extracted from hematite particles [Fig.2C] is added. This picture shows poor agreement with experimental data, as it neglects the force-free requirement.}
\label{Fig4}
\end{figure}

\section*{References}
\bibliography{BiblioAntoine.bib}

\begin{thebibliography}{10}
\expandafter\ifx\csname url\endcsname\relax
  \def\url#1{\texttt{#1}}\fi
\expandafter\ifx\csname urlprefix\endcsname\relax\def\urlprefix{URL }\fi
\providecommand{\bibinfo}[2]{#2}
\providecommand{\eprint}[2][]{\url{#2}}

\bibitem{Bechinger_Volpe-RevModPhys-2016}
\bibinfo{author}{Bechinger, C.} \emph{et~al.}
\newblock \bibinfo{title}{Active particles in complex and crowded
  environments}.
\newblock \emph{\bibinfo{journal}{Rev. Mod. Phys.}}
  \textbf{\bibinfo{volume}{88}}, \bibinfo{pages}{045006}
  (\bibinfo{year}{2016}).
\newblock
  \urlprefix\url{https://link.aps.org/doi/10.1103/RevModPhys.88.045006}.

\bibitem{Drescher_Goldstein-PNAS-2011}
\bibinfo{author}{Drescher, K.}, \bibinfo{author}{Dunkel, J.},
  \bibinfo{author}{Cisneros, L.~H.}, \bibinfo{author}{Ganguly, S.} \&
  \bibinfo{author}{Goldstein, R.~E.}
\newblock \bibinfo{title}{Fluid dynamics and noise in bacterial
  cell{\textendash}cell and cell{\textendash}surface scattering}.
\newblock \emph{\bibinfo{journal}{Proceedings of the National Academy of
  Sciences}} \textbf{\bibinfo{volume}{108}}, \bibinfo{pages}{10940--10945}
  (\bibinfo{year}{2011}).
\newblock \urlprefix\url{http://www.pnas.org/content/108/27/10940}.
\newblock \eprint{http://www.pnas.org/content/108/27/10940.full.pdf}.

\bibitem{Boniface_Tierno-NatComm-2024}
\bibinfo{author}{Boniface, D.}, \bibinfo{author}{Leyva, S.~G.},
  \bibinfo{author}{Pagonabarraga, I.} \& \bibinfo{author}{Tierno, P.}
\newblock \bibinfo{title}{Clustering induces switching between phoretic and
  osmotic propulsion in active colloidal rafts}.
\newblock \emph{\bibinfo{journal}{Nature Communications}}
  \textbf{\bibinfo{volume}{15}}, \bibinfo{pages}{5666} (\bibinfo{year}{2024}).
\newblock \urlprefix\url{https://doi.org/10.1038/s41467-024-49977-5}.

\bibitem{Blois_Dauchot-PRF-2019}
\bibinfo{author}{de~Blois, C.}, \bibinfo{author}{Reyssat, M.},
  \bibinfo{author}{Michelin, S.} \& \bibinfo{author}{Dauchot, O.}
\newblock \bibinfo{title}{Flow field around a confined active droplet}.
\newblock \emph{\bibinfo{journal}{Phys. Rev. Fluids}}
  \textbf{\bibinfo{volume}{4}}, \bibinfo{pages}{054001} (\bibinfo{year}{2019}).
\newblock
  \urlprefix\url{https://link.aps.org/doi/10.1103/PhysRevFluids.4.054001}.

\bibitem{Katuri_Sanchez-ScienceAdvances-2021}
\bibinfo{author}{Katuri, J.}, \bibinfo{author}{Uspal, W.~E.},
  \bibinfo{author}{Popescu, M.~N.} \& \bibinfo{author}{S?nchez, S.}
\newblock \bibinfo{title}{Inferring non-equilibrium interactions from tracer
  response near confined active janus particles}.
\newblock \emph{\bibinfo{journal}{Science Advances}}
  \textbf{\bibinfo{volume}{7}}, \bibinfo{pages}{eabd0719}
  (\bibinfo{year}{2021}).
\newblock
  \urlprefix\url{https://www.science.org/doi/abs/10.1126/sciadv.abd0719}.
\newblock \eprint{https://www.science.org/doi/pdf/10.1126/sciadv.abd0719}.

\bibitem{Campbell_Golestanian-NatureComm-2019}
\bibinfo{author}{Campbell, A.~I.}, \bibinfo{author}{Ebbens, S.~J.},
  \bibinfo{author}{Illien, P.} \& \bibinfo{author}{Golestanian, R.}
\newblock \bibinfo{title}{Experimental observation of flow fields around active
  janus spheres}.
\newblock \emph{\bibinfo{journal}{Nature Communications}}
  \textbf{\bibinfo{volume}{10}}, \bibinfo{pages}{3952} (\bibinfo{year}{2019}).
\newblock \urlprefix\url{https://doi.org/10.1038/s41467-019-11842-1}.

\bibitem{Simmchen_Sanchez-NatureComm-2016}
\bibinfo{author}{Simmchen, J.} \emph{et~al.}
\newblock \bibinfo{title}{Topographical pathways guide chemical microswimmers}.
\newblock \emph{\bibinfo{journal}{Nature Comm}} \textbf{\bibinfo{volume}{7}},
  \bibinfo{pages}{10598} (\bibinfo{year}{2016}).
\newblock \urlprefix\url{http://dx.doi.org/10.1038/ncomms10598}.

\bibitem{Bregulla_Cichos-JCP-2019}
\bibinfo{author}{Bregulla, A.~P.} \& \bibinfo{author}{Cichos, F.}
\newblock \bibinfo{title}{Flow fields around pinned self-thermophoretic
  microswimmers under confinement}.
\newblock \emph{\bibinfo{journal}{The Journal of Chemical Physics}}
  \textbf{\bibinfo{volume}{151}}, \bibinfo{pages}{044706}
  (\bibinfo{year}{2019}).
\newblock \urlprefix\url{https://doi.org/10.1063/1.5088131}.
\newblock \eprint{https://doi.org/10.1063/1.5088131}.

\bibitem{Singh_Mark_AdvancedMaterials-2017}
\bibinfo{author}{Singh, D.~P.}, \bibinfo{author}{Choudhury, U.},
  \bibinfo{author}{Fischer, P.} \& \bibinfo{author}{Mark, A.~G.}
\newblock \bibinfo{title}{Non-equilibrium assembly of light-activated colloidal
  mixtures}.
\newblock \emph{\bibinfo{journal}{Advanced Materials}}
  \textbf{\bibinfo{volume}{29}}, \bibinfo{pages}{1701328}
  (\bibinfo{year}{2017}).
\newblock
  \urlprefix\url{https://onlinelibrary.wiley.com/doi/abs/10.1002/adma.201701328}.
\newblock
  \eprint{https://onlinelibrary.wiley.com/doi/pdf/10.1002/adma.201701328}.

\bibitem{Yu_Fischer-ChemComm-2018}
\bibinfo{author}{Yu, T.} \emph{et~al.}
\newblock \bibinfo{title}{Chemical micromotors self-assemble and self-propel by
  spontaneous symmetry breaking}.
\newblock \emph{\bibinfo{journal}{Chem. Commun.}}
  \textbf{\bibinfo{volume}{54}}, \bibinfo{pages}{11933--11936}
  (\bibinfo{year}{2018}).
\newblock \urlprefix\url{http://dx.doi.org/10.1039/C8CC06467A}.

\bibitem{Xang_Simmchen-CondensedMatter-2019}
\bibinfo{author}{Wang, L.} \& \bibinfo{author}{Simmchen, J.}
\newblock \bibinfo{title}{Review: Interactions of active colloids with passive
  tracers}.
\newblock \emph{\bibinfo{journal}{Condensed Matter}}
  \textbf{\bibinfo{volume}{4}} (\bibinfo{year}{2019}).
\newblock \urlprefix\url{https://www.mdpi.com/2410-3896/4/3/78}.

\bibitem{Golestanian_Adjari-NewJournalofPhysics-2007}
\bibinfo{author}{Golestanian, R.}, \bibinfo{author}{Liverpool, T.~B.} \&
  \bibinfo{author}{Ajdari, A.}
\newblock \bibinfo{title}{Designing phoretic micro- and nano-swimmers}.
\newblock \emph{\bibinfo{journal}{New Journal of Physics}}
  \textbf{\bibinfo{volume}{9}}, \bibinfo{pages}{126} (\bibinfo{year}{2007}).
\newblock \urlprefix\url{http://stacks.iop.org/1367-2630/9/i=5/a=126}.

\bibitem{Karg_Hellweg-ChemPhysChem-2006}
\bibinfo{author}{Karg, M.}, \bibinfo{author}{Pastoriza-Santos, I.},
  \bibinfo{author}{Liz-Marz?n, L.~M.} \& \bibinfo{author}{Hellweg, T.}
\newblock \bibinfo{title}{A versatile approach for the preparation of
  thermosensitive pnipam core¡vshell microgels with nanoparticle cores}.
\newblock \emph{\bibinfo{journal}{ChemPhysChem}} \textbf{\bibinfo{volume}{7}},
  \bibinfo{pages}{2298--2301} (\bibinfo{year}{2006}).
\newblock
  \urlprefix\url{https://chemistry-europe.onlinelibrary.wiley.com/doi/abs/10.1002/cphc.200600483}.
\newblock
  \eprint{https://chemistry-europe.onlinelibrary.wiley.com/doi/pdf/10.1002/cphc.200600483}.

\bibitem{Youssef_Palacci-SoftMatter-2020}
\bibinfo{author}{Youssef, M.}, \bibinfo{author}{Morin, A.},
  \bibinfo{author}{Aubret, A.}, \bibinfo{author}{Sacanna, S.} \&
  \bibinfo{author}{Palacci, J.}
\newblock \bibinfo{title}{Rapid characterization of neutral polymer brush with
  a conventional zetameter and a variable pinch of salt}.
\newblock \emph{\bibinfo{journal}{Soft Matter}} \textbf{\bibinfo{volume}{16}},
  \bibinfo{pages}{4274--4282} (\bibinfo{year}{2020}).
\newblock \urlprefix\url{http://dx.doi.org/10.1039/C9SM01850F}.

\bibitem{Moyses_Grier-SoftMatter-2016}
\bibinfo{author}{Moyses, H.}, \bibinfo{author}{Palacci, J.},
  \bibinfo{author}{Sacanna, S.} \& \bibinfo{author}{Grier, D.~G.}
\newblock \bibinfo{title}{Trochoidal trajectories of self-propelled janus
  particles in a diverging laser beam}.
\newblock \emph{\bibinfo{journal}{Soft Matter}} \textbf{\bibinfo{volume}{12}},
  \bibinfo{pages}{6357--6364} (\bibinfo{year}{2016}).
\newblock \urlprefix\url{http://dx.doi.org/10.1039/C6SM01163B}.

\bibitem{Spagnolie_Lauga-JFM-2012}
\bibinfo{author}{Spagnolie, S.~E.} \& \bibinfo{author}{Lauga, E.}
\newblock \bibinfo{title}{Hydrodynamics of self propulsion near a boundary
  predictions and accuracy of far-field approximations}.
\newblock \emph{\bibinfo{journal}{Journal of Fluid Mechanics}}
  \textbf{\bibinfo{volume}{700}}, \bibinfo{pages}{105} (\bibinfo{year}{2012}).

\bibitem{Reigh_Kapral-SoftMatter-2015}
\bibinfo{author}{Reigh, S.~Y.} \& \bibinfo{author}{Kapral, R.}
\newblock \bibinfo{title}{Catalytic dimer nanomotors: continuum theory and
  microscopic dynamics}.
\newblock \emph{\bibinfo{journal}{Soft Matter}} \textbf{\bibinfo{volume}{11}},
  \bibinfo{pages}{3149--3158} (\bibinfo{year}{2015}).
\newblock \urlprefix\url{http://dx.doi.org/10.1039/C4SM02857K}.

\bibitem{Michelin_Lauga-ScientificReports-2017}
\bibinfo{author}{Michelin, S.} \& \bibinfo{author}{Lauga, E.}
\newblock \bibinfo{title}{Geometric tuning of self-propulsion for janus
  catalytic particles}.
\newblock \emph{\bibinfo{journal}{Scientific Reports}}
  \textbf{\bibinfo{volume}{7}}, \bibinfo{pages}{42264} (\bibinfo{year}{2017}).
\newblock \urlprefix\url{https://doi.org/10.1038/srep42264}.

\bibitem{Ishimoto_Gaffney-PRE-2013}
\bibinfo{author}{Ishimoto, K.} \& \bibinfo{author}{Gaffney, E.~A.}
\newblock \bibinfo{title}{Squirmer dynamics near a boundary}.
\newblock \emph{\bibinfo{journal}{Phys. Rev. E}} \textbf{\bibinfo{volume}{88}},
  \bibinfo{pages}{062702} (\bibinfo{year}{2013}).
\newblock \urlprefix\url{https://link.aps.org/doi/10.1103/PhysRevE.88.062702}.

\bibitem{Wagner_Ripoll-EPJE-2021}
\bibinfo{author}{Wagner, M.}, \bibinfo{author}{Roca-Bonet, S.} \&
  \bibinfo{author}{Ripoll, M.}
\newblock \bibinfo{title}{Collective behavior of thermophoretic dimeric active
  colloids in three-dimensional bulk}.
\newblock \emph{\bibinfo{journal}{The European Physical Journal E}}
  \textbf{\bibinfo{volume}{44}}, \bibinfo{pages}{43} (\bibinfo{year}{2021}).
\newblock \urlprefix\url{https://doi.org/10.1140/epje/s10189-021-00043-8}.

\bibitem{Liebchen_Lowen-JCP-2019}
\bibinfo{author}{Liebchen, B.} \& \bibinfo{author}{L?wen, H.}
\newblock \bibinfo{title}{Which interactions dominate in active colloids?}
\newblock \emph{\bibinfo{journal}{J. Chem. Phys.}}
  \textbf{\bibinfo{volume}{150}}, \bibinfo{pages}{061102}
  (\bibinfo{year}{2019}).
\newblock \urlprefix\url{https://doi.org/10.1063/1.5082284}.

\bibitem{Liebchen_Cates-PRL-2017}
\bibinfo{author}{Liebchen, B.}, \bibinfo{author}{Marenduzzo, D.} \&
  \bibinfo{author}{Cates, M.~E.}
\newblock \bibinfo{title}{Phoretic interactions generically induce dynamic
  clusters and wave patterns in active colloids}.
\newblock \emph{\bibinfo{journal}{Phys. Rev. Lett.}}
  \textbf{\bibinfo{volume}{118}}, \bibinfo{pages}{268001}
  (\bibinfo{year}{2017}).
\newblock
  \urlprefix\url{https://link.aps.org/doi/10.1103/PhysRevLett.118.268001}.

\bibitem{Spagolie_Lauga-SoftMatter-2015}
\bibinfo{author}{Spagnolie, S.~E.}, \bibinfo{author}{Moreno-Flores, G.~R.},
  \bibinfo{author}{Bartolo, D.} \& \bibinfo{author}{Lauga, E.}
\newblock \bibinfo{title}{Geometric capture and escape of a microswimmer
  colliding with an obstacle}.
\newblock \emph{\bibinfo{journal}{Soft Matter}} \textbf{\bibinfo{volume}{11}},
  \bibinfo{pages}{3396--3411} (\bibinfo{year}{2015}).
\newblock \urlprefix\url{http://dx.doi.org/10.1039/C4SM02785J}.

\bibitem{Ibrahim_Liverpool-EL_2015}
\bibinfo{author}{Ibrahim, Y.} \& \bibinfo{author}{Liverpool, T.~B.}
\newblock \bibinfo{title}{The dynamics of a self-phoretic janus swimmer near a
  wall}.
\newblock \emph{\bibinfo{journal}{Europhysics Letters}}
  \textbf{\bibinfo{volume}{111}}, \bibinfo{pages}{48008}
  (\bibinfo{year}{2015}).
\newblock \urlprefix\url{https://dx.doi.org/10.1209/0295-5075/111/48008}.

\bibitem{Popescu_Dietrich-EPJE-2018}
\bibinfo{author}{Popescu, M.~N.}, \bibinfo{author}{Uspal, W.~E.},
  \bibinfo{author}{Eskandari, Z.}, \bibinfo{author}{Tasinkevych, M.} \&
  \bibinfo{author}{Dietrich, S.}
\newblock \bibinfo{title}{Effective squirmer models for self-phoretic
  chemically active spherical colloids}.
\newblock \emph{\bibinfo{journal}{The European Physical Journal E}}
  \textbf{\bibinfo{volume}{41}}, \bibinfo{pages}{145} (\bibinfo{year}{2018}).
\newblock \urlprefix\url{https://doi.org/10.1140/epje/i2018-11753-1}.

\bibitem{Bayati_Najafi-SoftMatter-2019}
\bibinfo{author}{Bayati, P.}, \bibinfo{author}{Popescu, M.~N.},
  \bibinfo{author}{Uspal, W.~E.}, \bibinfo{author}{Dietrich, S.} \&
  \bibinfo{author}{Najafi, A.}
\newblock \bibinfo{title}{Dynamics near planar walls for various model
  self-phoretic particles}.
\newblock \emph{\bibinfo{journal}{Soft Matter}} \textbf{\bibinfo{volume}{15}},
  \bibinfo{pages}{5644--5672} (\bibinfo{year}{2019}).
\newblock \urlprefix\url{http://dx.doi.org/10.1039/C9SM00488B}.

\bibitem{Uspal_Tasinkevych-SoftMatter-2015}
\bibinfo{author}{Uspal, W.~E.}, \bibinfo{author}{Popescu, M.~N.},
  \bibinfo{author}{Dietrich, S.} \& \bibinfo{author}{Tasinkevych, M.}
\newblock \bibinfo{title}{Self-propulsion of a catalytically active particle
  near a planar wall: from reflection to sliding and hovering}.
\newblock \emph{\bibinfo{journal}{Soft Matter}} \textbf{\bibinfo{volume}{11}},
  \bibinfo{pages}{434--438} (\bibinfo{year}{2015}).
\newblock \urlprefix\url{http://dx.doi.org/10.1039/C4SM02317J}.

\bibitem{Shen_Lintuvuori-EPJE-2018}
\bibinfo{author}{Shen, Z.}, \bibinfo{author}{W?rger, A.} \&
  \bibinfo{author}{Lintuvuori, J.~S.}
\newblock \bibinfo{title}{Hydrodynamic interaction of a self-propelling
  particle with a wall}.
\newblock \emph{\bibinfo{journal}{The European Physical Journal E}}
  \textbf{\bibinfo{volume}{41}}, \bibinfo{pages}{39} (\bibinfo{year}{2018}).
\newblock \urlprefix\url{https://doi.org/10.1140/epje/i2018-11649-0}.

\bibitem{Liebchen_Mukhopadhyay-JPCM-2022}
\bibinfo{author}{Liebchen, B.} \& \bibinfo{author}{Mukhopadhyay, A.~K.}
\newblock \bibinfo{title}{Interactions in active colloids}.
\newblock \emph{\bibinfo{journal}{Journal of Physics: Condensed Matter}}
  \textbf{\bibinfo{volume}{34}}, \bibinfo{pages}{083002}
  (\bibinfo{year}{2021}).
\newblock \urlprefix\url{https://dx.doi.org/10.1088/1361-648X/ac3a86}.

\bibitem{Ketzetzi_Kraft-NatureComm-2022}
\bibinfo{author}{Ketzetzi, S.}, \bibinfo{author}{Rinaldin, M.},
  \bibinfo{author}{Dr?ge, P.}, \bibinfo{author}{Graaf, J.~d.} \&
  \bibinfo{author}{Kraft, D.~J.}
\newblock \bibinfo{title}{Activity-induced interactions and cooperation of
  artificial microswimmers in one-dimensional environments}.
\newblock \emph{\bibinfo{journal}{Nature Communications}}
  \textbf{\bibinfo{volume}{13}}, \bibinfo{pages}{1772} (\bibinfo{year}{2022}).
\newblock \urlprefix\url{https://doi.org/10.1038/s41467-022-29430-1}.

\bibitem{Das_Ebbens-NatureComm-2015}
\bibinfo{author}{Das, S.} \emph{et~al.}
\newblock \bibinfo{title}{Boundaries can steer active janus spheres}.
\newblock \emph{\bibinfo{journal}{Nature Comm}} \textbf{\bibinfo{volume}{6}},
  \bibinfo{pages}{8999} (\bibinfo{year}{2015}).
\newblock \urlprefix\url{http://dx.doi.org/10.1038/ncomms9999}.

\bibitem{RojasPerez_Michelin-JFM-2021}
\bibinfo{author}{Rojas-P?rez, F.}, \bibinfo{author}{Delmotte, B.} \&
  \bibinfo{author}{Michelin, S.}
\newblock \bibinfo{title}{Hydrochemical interactions of phoretic particles: a
  regularized multipole framework}.
\newblock \emph{\bibinfo{journal}{Journal of Fluid Mechanics}}
  \textbf{\bibinfo{volume}{919}}, \bibinfo{pages}{A22} (\bibinfo{year}{2021}).

\bibitem{Lauga_Michelin-PRL-2016}
\bibinfo{author}{Lauga, E.} \& \bibinfo{author}{Michelin, S.}
\newblock \bibinfo{title}{Stresslets induced by active swimmers}.
\newblock \emph{\bibinfo{journal}{Phys. Rev. Lett.}}
  \textbf{\bibinfo{volume}{117}}, \bibinfo{pages}{148001}
  (\bibinfo{year}{2016}).
\newblock
  \urlprefix\url{https://link.aps.org/doi/10.1103/PhysRevLett.117.148001}.

\bibitem{Drescher_Tuval-PRL-2010}
\bibinfo{author}{Drescher, K.}, \bibinfo{author}{Goldstein, R.~E.},
  \bibinfo{author}{Michel, N.}, \bibinfo{author}{Polin, M.} \&
  \bibinfo{author}{Tuval, I.}
\newblock \bibinfo{title}{Direct measurement of the flow field around swimming
  microorganisms}.
\newblock \emph{\bibinfo{journal}{Phys. Rev. Lett.}}
  \textbf{\bibinfo{volume}{105}}, \bibinfo{pages}{168101}
  (\bibinfo{year}{2010}).
\newblock
  \urlprefix\url{https://link.aps.org/doi/10.1103/PhysRevLett.105.168101}.

\end{thebibliography}
\bibliographystyle{naturemag}
\subsection*{Conflict of interest}
 The authors declare no conflicts of interests.
\subsection*{Acknowledgments} We thank M. Perrin and A. Allard for enlightening discussions . This research was funded in whole or in part by the Austrian Science Fund (FWF) [https://doi.org/10.55776/P35206]. This project has received funding from the European Union’s Horizon 2020 research and innovation programme under the Marie Sklodowska Curie grant agreement No 886024.
\subsection*{Supplementary information} Supplementary information is available in the online version of the paper or on request.
\subsection*{Correspondence} Correspondence and requests for materials should be addressed to AA.~(email: antoine.aubret@u-bordeaux.fr).

\end{document}